\documentclass[twocolumn,aps,xamsmath,amssymb,prl,nofootinbib]{revtex4-1}
\usepackage{graphicx}
\usepackage{epstopdf}
\usepackage{dcolumn}
\usepackage{bm}
\usepackage{hyperref}
\usepackage{color}
\usepackage{amsmath}
\usepackage{url}

\newcommand{\vev}[1]{ \left\langle {#1} \right\rangle }

\def\Msusy{m_{\rm stop}}

\begin{document}

\title{Asymptotically Free Natural SUSY Twin Higgs}

\author{Marcin Badziak}
\affiliation{Institute of Theoretical Physics, Faculty of Physics, University of Warsaw, ul.~Pasteura 5, PL--02--093 Warsaw, Poland}
\affiliation{Department of Physics, University of California, Berkeley, California 94720, USA}
\affiliation{Theoretical Physics Group, Lawrence Berkeley National Laboratory, Berkeley, California 94720, USA}
\author{Keisuke Harigaya}
\affiliation{Department of Physics, University of California, Berkeley, California 94720, USA}
\affiliation{Theoretical Physics Group, Lawrence Berkeley National Laboratory, Berkeley, California 94720, USA}

\begin{abstract}
Twin Higgs (TH) models explain the absence of new colored particles responsible for natural electroweak symmetry
breaking (EWSB). 
All known ultraviolet completions of TH models require some non-perturbative dynamics below the Planck scale.
We propose a supersymmetric model in which the TH mechanism is introduced by a new asymptotically free gauge interaction.
The model features natural EWSB for squarks and gluino heavier than 2 TeV even if supersymmetry breaking is mediated around the Planck scale, and has
interesting flavor phenomenology including the top quark decay into the Higgs and the up quark which
may be discovered at the LHC.
\end{abstract}

\maketitle

{\it Introduction.}---%
Models of natural electroweak symmetry breaking (EWSB), such as supersymmetric (SUSY) models~\cite{MaianiLecture,Veltman:1980mj,Witten:1981nf,Kaul:1981wp} and composite Higgs models~\cite{Kaplan:1983fs,Kaplan:1983sm},  generically predict new
light
colored particles, called top partners, so that the quantum correction to the Higgs mass is suppressed. Null results of
the LHC searches, however, show that new colored particles are heavy, which calls for fine-tuning of the parameters of the theories; this is known as
the little hierarchy problem.
In light of this fact the idea that the light top partners are not charged under the Standard Model (SM) $SU(3)_c$ gauge group has become increasingly
attractive.
Twin Higgs (TH) models~\cite{Chacko:2005pe} are one of the most studied realizations of the idea.

A crucial ingredient of TH models is an approximate global $SU(4)$ symmetry under which the SM Higgs and its mirror (or twin) partner transform as a
fundamental
representation. The Higgs boson is a
pseudo-Nambu-Goldstone boson
associated with the spontaneous breakdown of the $SU(4)$ symmetry.
The $SU(4)$ symmetry of the Higgs mass term emerges from a $\mathbb{Z}_2$ symmetry exchanging the SM fields with their
mirror counterparts. The light top partners are then charged under the mirror gauge group rather than the SM one.
Standard lore says that ultraviolet (UV) completion of TH models involves some
non-perturbative
dynamics. This is because the quality of the $SU(4)$ symmetry requires a large $SU(4)$ invariant quartic term which points to UV
completions based on Composite Higgs models \cite{Batra:2008jy,Geller:2014kta,Barbieri:2015lqa,Low:2015nqa}. SUSY UV completions of the TH model also
exist \cite{Falkowski:2006qq,Chang:2006ra,Craig:2013fga,Katz:2016wtw,Badziak:2017syq,Badziak:2017kjk}. Acceptable tuning of the electroweak (EW) scale at the level
of $5-10\%$ can be, however, obtained only with a low Landau pole scale, which requires UV completion by some strong dynamics.
SUSY models that are able to keep the tuning at the level of $5-10\%$ without resorting to the TH mechanism also require
a low cut-off scale~\cite{Dimopoulos:2014aua}.

In this Letter we propose a SUSY Twin Higgs model with an asymptotically free $SU(4)$ invariant quartic coupling. The model remains perturbative up to
around the Planck scale, and does not require any further UV completion below the energy scale of gravity. As a result the yukawa couplings of the SM particles are given by  renormalizable interactions.

{\it Setup.}---%
It was proposed in~\cite{Badziak:2017syq} that an $SU(4)$ invariant quartic coupling may be obtained from a D term potential of a new $U(1)_X$ gauge
symmetry under which the SM
and mirror Higgses are charged.
The model suffers from a low Landau pole scale of the $U(1)_X$ gauge interaction. A model with a non-Abelian
$SU(2)_X$ gauge symmetry was proposed in~\cite{Badziak:2017kjk}, so that the Landau pole scale is far above the TeV scale. Still the gauge interaction
is
asymptotically non-free. In order for the gauge interaction to be perturbative up to a high energy scale of $10^{16\mathchar`-18}$ GeV, the $SU(4)$
invariant quartic coupling at the TeV scale must be small, and the TH mechanism does not work perfectly well; fine-tuning of order one percent is
required to obtain a correct EWSB scale.

In this Letter, we present an extension of the model such that the new gauge interaction is asymptotically free.
In the model presented in~\cite{Badziak:2017kjk}, the new gauge symmetry $SU(2)_X$ is assumed to be $\mathbb{Z}_2$  neutral, and mirror particles are charged
under $SU(2)_X$.
We instead assume that $SU(2)_X$ has a mirror partner $SU(2)_X'$, under which mirror particles are charged.
As a result the number of $SU(2)_X$ charged fields is reduced, so that the $SU(2)_X$ gauge interaction is asymptotically free.
A similar group structure in a non-Twin SUSY model was introduced in~\cite{Batra:2003nj} to achieve asymptotically free gauge theory. 

The charged matter content of the model is shown in Table~\ref{tab:matter}.
The up-type SM and mirror Higgses are embedded into ${\cal H}$ and ${\cal H}'$, respectively.
The resultant D term potentials of the gauge symmetries are not $SU(4)$ invariant. Once $SU(2)_X \times SU(2)_X'$ symmetry is broken down to a
diagonal subgroup $SU(2)_D$ 
by a non-zero vacuum expectation value (VEV) of a bi-fundamental $\Sigma$, both the SM and mirror Higgses are fundamental representation of
$SU(2)_D$,
and the D term potential is approximately $SU(4)$ invariant below the symmetry breaking scale.
The $SU(2)_D$ symmetry is completely broken down by the VEVs of $S$, $\bar{S}$, $S'$, $\bar{S}'$.

The right-handed top quark is embedded into $\bar{Q}_R$, so that a large enough top yukawa coupling is obtained via the superpotential term $W \sim 
{\cal H} \bar{Q}_R Q_3$, where $Q_i$ is the $i\mathchar`-$th generation of left-handed quarks.
The right-handed up quark is also embedded into $\bar{Q}_R$.
The VEV of $\phi_u$ gives a mass to the charm quark via $W \sim \phi_u \bar{u}_2 Q_2$.
We assume that yukawa couplings ${\cal H} \bar{Q}_R Q_{1,2}$ and $\phi_u \bar{u}_2 Q_{1,3}$ are small so that tree level flavor changing neutral currents (FCNCs) are suppressed.
$H_d$
gives masses to down-type quarks and charged leptons via $W \sim H_d \bar{d} Q + H_d L \bar{e}$.
We assume that the yukawa couplings involving $\phi_{d,1,2}$ are suppressed; otherwise large FCNCs are induced.
$H_d$ and $\phi_{d,1,2}$  are the
mass partners of ${\cal H}$ and $\phi_u$,
for details see~\cite{Badziak:2017kjk}.
Due to the $SU(2)_X$ invariance, after ${\cal H}$ and $\phi_u$ obtain their VEVs, one linear combination of the two components in $\bar{Q}_R$ remains
massless at the tree level. 
The one loop quantum correction with a charged wino, charged higgsinos in ${\cal H}$ and down-type left-handed squarks inside the loop generates the
up-quark mass.
The mass of the higgsinos in ${\cal H}$ is given by the $SU(2)_X$ symmetry breaking and hence the loop mediate the breaking.
The field $\bar{E}$ cancels the anomaly of $U(1)_Y\mathchar`-SU(2)_X^2$, while $E_{1,2}$ cancels that of $U(1)_Y^3$. The charged lepton is in general the mixture of $\bar{E}$, $E$, $\bar{e}$ and the charged component of $L$ due to possible mixing $W \sim \bar{e} E$.

\begin{figure}
 \includegraphics[width=0.48\textwidth]{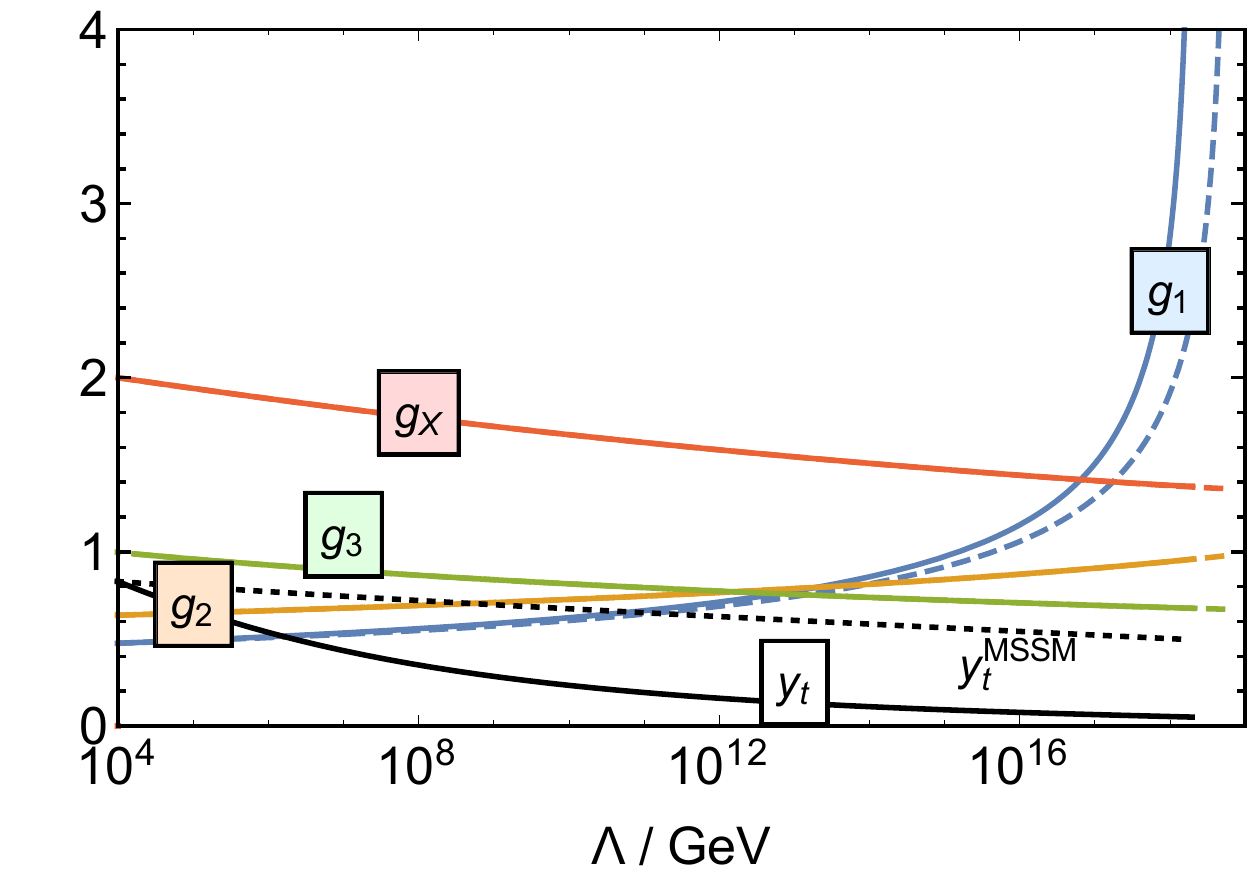}
 \caption{RG running of $g_X$ (red), $g_1$ (blue), $g_2$ (yellow), $g_3$ (green) and the top yukawa coupling $y_t$ (black) for $m_X=10$~TeV, $m_{\rm
stop}=2$~TeV, $g_X(m_X)=2$ and $\tan\beta=3$. Solid lines correspond to the case where all states beyond the Minimal Supersymmetric Standard Model
(MSSM) have masses around $m_X$. Dashed lines assume $M_{E_1}=10^7$~GeV, $M_{E_2}=10^9$~GeV, see text for
details.
Dotted black line corresponds to the running
of $y_t$
in the MSSM.
}
\label{fig:running}
\end{figure}

The number of fundamental representations of $SU(2)_X$ is $10$. Thus the $SU(2)_X$ gauge interaction is asymptotically free, unless
$g_X\gtrsim3.2$ for which two-loop correction changes the sign of the beta function for $g_X$.
In Fig.~\ref{fig:running}, we show the renormalization group (RG) running of the gauge coupling constants and the top yukawa coupling, where we use
the NSVZ beta
function~\cite{Novikov:1983uc} with the anomalous
dimension evaluated at the one-loop level. This explicitly confirms asymptotically-free behavior of the new interaction. 
Here and hereafter, we approximate the RG running above the $SU(2)_D$ symmetry breaking scale by that of the $SU(2)_X \times SU(2)_X'$ symmetric
theory.
This is a good approximation as long as the $SU(2)_X \times SU(2)_X'$ breaking scale is within the same order of magnitude as the $SU(2)_D$ breaking
scale.

The model possesses many new states with non-zero hypercharge which make the appearance of the Landau pole for the hypercharge much
lower than in the SM. Nevertheless, this Landau pole appears around $10^{18}$~GeV, as seen from Fig.~\ref{fig:running}, which is rather close to the
Planck scale. The Landau pole scale is pushed up if some of new states are much heavier than the TeV scale.
Actually we can give a large Dirac mass term $M_{E_1}\bar{e}_1 E_1 + M_{E_2}\bar{e}_2 E_2 $. After integrating them out, the electron and muon masses are given by a dimension-5 term $ W \sim (S \phi_d \bar{E}L_i + \bar{S} \phi_d \bar{E}L_i) / M_{E_1} $.
For ${\cal{O}}(1)$ coupling of $W \sim S \bar{E}E + \phi_d \bar{E} L$, the Dirac masses may be as large as $M_{E_1}\approx10^7$~GeV,
$M_{E_2}\approx10^9$~GeV. The RG running in such a case is also shown in Fig.~\ref{fig:running}.

\begin{table}[htp]
\caption{The charged matter content of the model.}
\begin{center}
\begin{tabular}{|c|c|c|c|c|}
\hline
                           & $SU(2)_X$ & $SU(2)_X'$ &$3\mathchar`-2\mathchar`-1$&$3'\mathchar`-2'\mathchar`-1'$ \\ \hline
${\cal H}$            & ${\bf 2}$      &                     &  $({\bf 1},{\bf 2},1/2)$            &                                                  \\
${\cal H}'$           &                    & ${\bf 2}$      &                                               &  $({\bf 1},{\bf 2},1/2)$               \\
$\Sigma$              & ${\bf 2}$      &${\bf 2}$       &                                               &                                            
    \\
$S$                     & ${\bf 2}$      &                    &                                               &                                                  \\
$\bar{S}$             & ${\bf 2}$     &                    &                                               &                                                  \\
$S'$                     &                   & ${\bf 2}$      &                                               &                                                  \\
$\bar{S}'$            &                   & ${\bf 2}$      &                                               &                                                  \\
$\bar{Q}_R$       & ${\bf 2}$      &                     &  $({\bf \bar{3}},{\bf 1},-2/3)$            &                                                  \\
$\bar{Q}_R'$       &                    & ${\bf 2}$      &                                               &  $({\bf 3},{\bf 1},-2/3)$               \\
$\bar{E}$             & ${\bf 2}$     &                     &  $({\bf 1},{\bf 1},1)$            &                                                  \\
$\bar{E}'$            &                    & ${\bf 2}$      &                                           &  $({\bf 1},{\bf 1},1)$  
                \\
$E_{1,2}$             &                   &                     &  $({\bf 1},{\bf 1},-1)$            &                                                  \\
$E_{1,2}'$            &                   &                     &                                            &  $({\bf 1},{\bf 1},-1)$                \\
$\phi_u$              &                    &                     &  $({\bf 1},{\bf 2},1/2)$            &                                                \\
$\phi_u'$              &                    &                     &                                             &  $({\bf 1},{\bf 2},1/2)$              \\
$H_d, \phi_{d,1,2}$   &                    &                     &  $({\bf 1},{\bf 2},-1/2)$            &                                           
  \\
$H_d', \phi_{d,1,2}'$   &                    &                     &                                              &  $({\bf 1},{\bf 2},-1/2)$       
   \\
$Q_{1,2,3}$         &                    &                     &  $({\bf 3},{\bf 2},1/6)$            &                                                \\
$\bar{u}_{2}$         &                    &                   &  $({\bf \bar{3}},{\bf 1},-2/3)$    &                                               
\\
$\bar{e}_{1,2,3}$         &                    &             &  $({\bf 1},{\bf 1},1)$                &                                                \\
$\bar{d}_{1,2,3}$         &                    &             &  $({\bf \bar{3}},{\bf 1},1/3)$      &                                                \\
$L_{1,2,3}$         &                    &             &  $({\bf 1},{\bf 2},-1/2)$            &                                                \\
$Q_{1,2,3}'$         &                    &                     &                                               &  $({\bf 3},{\bf 2},1/6)$             \\
$\bar{u}_{2}'$         &                    &                   &                                                &  $({\bf \bar{3}},{\bf 1},-2/3)$   
\\
$\bar{e}_{1,2,3}'$         &                    &             &                                                &  $({\bf 1},{\bf 1},1)$                \\
$\bar{d}_{1,2,3}'$         &                    &             &                                                 &  $({\bf \bar{3}},{\bf 1},1/3)$     \\
$\bar{L}_{1,2,3}'$         &                    &             &                                                &  $({\bf 1},{\bf 2},-1/2)$             \\ \hline
\end{tabular}
\end{center}
\label{tab:matter}
\end{table}%

Let us evaluate the magnitude of the $SU(4)$ invariant coupling. We assume that $\Sigma$ obtains its VEV in a SUSY way, e.g.~by a
superpotential $W = Y (\Sigma^2 - v_\Sigma^2)$, and that $\vev{\Sigma}$ is much larger than the TeV scale, say few tens of TeV. Then below the scale
$v_\Sigma$ the theory is well-described by a SUSY theory with an $SU(2)_D$ gauge symmetry. The symmetry breaking of $SU(2)_D$ should involve
SUSY breaking effect, so that the D term potential of $SU(2)_D$ does not decouple after the symmetry breaking. We introduce the superpotential
\begin{align}
W = \kappa \Xi (S\bar{S}- M^2) + \kappa \Xi' (S'\bar{S}'- M^2)
\end{align}
and soft masses,
\begin{align}
V_{\rm soft} = m_S^2 (|S|^2 + |\bar{S}|^2 +  |S'|^2 + |\bar{S}'|^2 ).
\end{align}
Here we assume that the soft masses of $S$ and $\bar{S}$ are the same. Otherwise, the asymmetric VEVs of $S$ and $\bar{S}$ give a large soft mass to
the Higgs doublet through the D term potential of $SU(2)_D$. Assuming that all Higgses apart from the SM-like and twin Higgs 
are heavy, negligible VEV of $\phi_u$ and integrating out $S$ fields, the $SU(4)$ invariant quartic coupling of the SM Higgs $H$ and the mirror
Higgs $H'$ is given by
\begin{align}
V = \frac{g_X^2}{8}{\rm sin}^4 \beta (1- \epsilon^2) (|H|^2 + |H'|^2 )^2, \nonumber \\ 
\epsilon^2 \equiv \frac{m_X^2}{2m_S^2 + m_X^2},
\end{align}
where ${\rm tan}\beta$ is the ratio of the up-type Higgs component to the down-type Higgs component in $H$.

{\it Natural electroweak symmetry breaking.}---%
Asymptotic freedom of the new gauge interactions allows the $SU(4)$ invariant coupling of ${\cal{O}}(1)$.
The tuning of the EW scale
arising from heaviness of
higgsino, stops and gluino, may be suppressed even by a factor of ${\cal{O}}(10)$ by means of the TH mechanism alone.
Moreover, large $g_X$ strongly suppresses the top yukawa coupling at high energy scales, as seen from Fig.~\ref{fig:running}, which results in
additional suppression of
the correction to the Higgs mass parameter from stops and gluino. However, for very large values of $g_X$, close to the perturbativity bound for the
$SU(2)_X$ interaction, the tuning of the EW scale is dominated by a finite threshold correction from the gauge bosons of the new interaction:
\begin{equation}
\label{deltamHu_X}
 \left(\delta m_{H_u}^2\right)_{X}= 3 \frac{g_X^2}{64 \pi^2} m_X^2 \ln\left(\epsilon^{-2}\right) \,.
\end{equation}
For large values of $g_X$, that we are most interested in, the strongest lower mass limit on the new gauge boson mass of $m_X\gtrsim g_X\times 4$~TeV 
originates from the mixing between the $Z$ boson and the $SU(2)_D$ gauge bosons which breaks custodial symmetry, see~\cite{Badziak:2017kjk} for a
detailed derivation of this bound using the EW precision observables. The threshold correction in Eq.~\eqref{deltamHu_X} is smaller for larger $\epsilon$
which leads also to smaller $SU(4)$ invariant coupling; some intermediate value of $\epsilon$ is optimal from the point of view of tuning of the EW
scale. Not too small $\epsilon$, i.e.~not too heavy $S$ fields, is also preferred too avoid a large two-loop correction to $m_{H_u}^2$ proportional
to $g_X^4m_S^2$. 

In order to quantify the tuning we use the measure~\cite{Craig:2013fga}
\begin{equation}
\label{eq:Delta_v}
\Delta_v \equiv   \Delta_f \times \Delta_{v/f},
\end{equation}
where the tuning in percent is $100\%/\Delta_v$ and
\begin{align}
\label{eq:Delta_vf}
&\Delta_{v/f} = \frac{1}{2} \left( \frac{f^2}{v^2} -2\right), \\
\label{eq:Delta_f}
&\Delta_f =  {\rm max}_i \left( |\frac{\partial{\rm ln} f^2}{\partial{\rm ln} x_i(\Lambda)}|, 1 \right) .
\end{align}
Here $\vev{H} \equiv v$, $\vev{H'}\equiv v'$, and $f \equiv \sqrt{v^2 + v^{'2}}$ is the decay constant of the spontaneous $SU(4)$ breaking.
$\Delta_{v/f}$ measures the tuning to obtain $v < f$
via explicit soft $\mathbb{Z}_2$ symmetry breaking which is required by the Higgs coupling measurements~\cite{Higgscomb}, implying
$f\gtrsim2.3v$~\cite{Buttazzo:2015bka}. In our numerical analysis we fix $f=3v$. $\Delta_f$ measures the tuning to obtain the scale $f$ from the soft
SUSY breaking which is
analogous to the fine-tuning to obtain the EW scale from the soft SUSY breaking in the MSSM.
$x_i(\Lambda)$ are the parameters of the theory evaluated at the mediation scale of the SUSY breaking $\Lambda$ including $m_{H_u}^2$, $m_{Q_3}^2$,
$m_{\bar{u}_3}^2$, $M_1^2$, $M_2^2$, $M_3^2$, $\mu^2$, $m_S^2$ and $m_\Xi^2$,
where $m_\Xi^2$ is the soft mass of $\Xi$.
In the following numerical analysis we assume $m_S^2=0$ at the
mediation scale and a value of $m_\Xi^2$ such that $m_S=m_X$ at the $SU(2)_D$ breaking scale, corresponding to $\epsilon^2=1/3$, is generated via the
RG running with $\kappa=0.2$ at the mediation scale, see~\cite{Badziak:2017kjk} for more details of the calculation of $\Delta_v$.

\begin{figure}
 \includegraphics[width=0.48\textwidth]{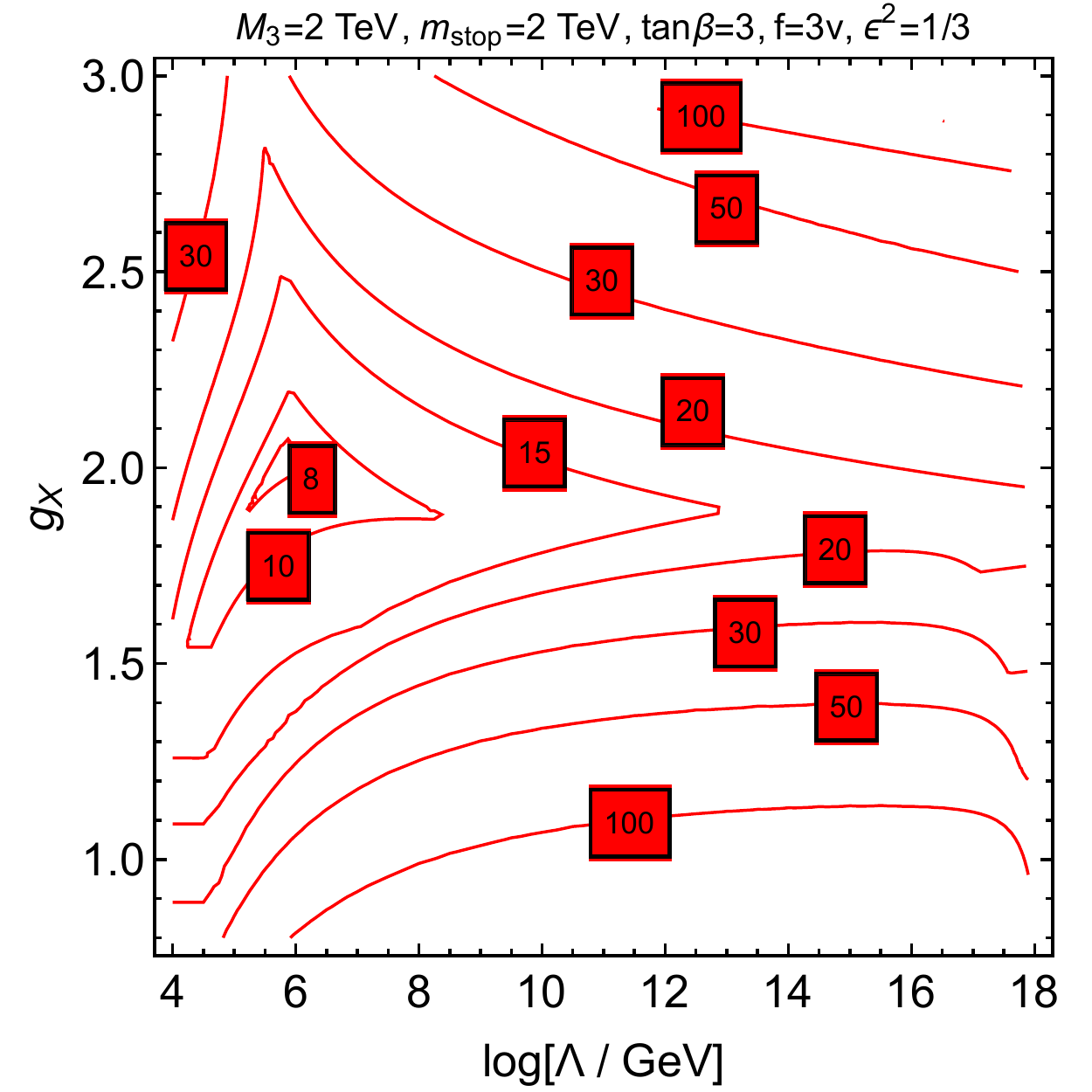}
 \caption{Fine-tuning $\Delta_v$ of the model in the plane $\Lambda$-$g_X$ for  $\Msusy=2$~TeV, $\tan\beta=3$, $f=3v$, $\mu=500$~GeV,
$M_1=M_2=200$~GeV and the soft gluino mass term $M_3=2$ TeV. We fix $\epsilon^2=1/3$ which corresponds to $m_S=m_X$.
}
\label{fig:ft}
\end{figure}

The tuning in the plane $\Lambda$-$g_X$ for $\Msusy=M_3=2$~TeV at the TeV scale is shown in Fig.~\ref{fig:ft}. The tuning does not depend strongly on $\tan\beta$ so we
fix $\tan\beta=3$ which leads to the Higgs mass consistent with the Higgs mass measurement within theoretical uncertainties,
see~\cite{Badziak:2017syq} for a more detailed discussion of the Higgs mass in SUSY TH models.  Wee see that the tuning decreases with
increasing $g_X$ as a consequence of the TH mechanism as long as $g_X\lesssim2$. For larger $g_X$ the tuning becomes dominated by the threshold
correction in Eq.~\eqref{deltamHu_X} and the two-loop correction from the soft masses of $S$ fields, so further increasing $g_X$ worsens the tuning. For the optimal value of
$g_X\approx2$ the tuning is only at the level
of $5-10\%$ even for very large mediation scales. This allows to employ gravitational interactions as a source of SUSY breaking mediation without
excessive fine-tuning, 
in contrast to the MSSM and previously proposed SUSY TH models.

The above discussion of tuning, similarly to all previous papers on SUSY TH models, assumed the soft stop masses at the low scale as an input without
paying attention to the question of what kind of SUSY breaking mechanism can realize such a spectrum. Since in this model TH mechanism is at work also
for high mediation scales, we calculate the spectrum using simple UV boundary conditions.
We assume a universal soft scalar masses $m_0$ for the SM charged fields at the mediation scale, which explains the smallness of the flavor violation from SUSY particles.
$m_S^2$ and $m_{\Xi}^2$ are determined in the same way as before.
We fix all soft trilinear terms $A_0=0$ at the mediation scale.
On the other hand, there is no well-motivated choice for gaugino masses since in this model the gauge couplings do not unify.
Thus, similarly as before we take gaugino masses at the low scale as input. We fix $M_1=M_2=200$~GeV and
vary $M_3$.

\begin{figure}
 \includegraphics[width=0.48\textwidth]{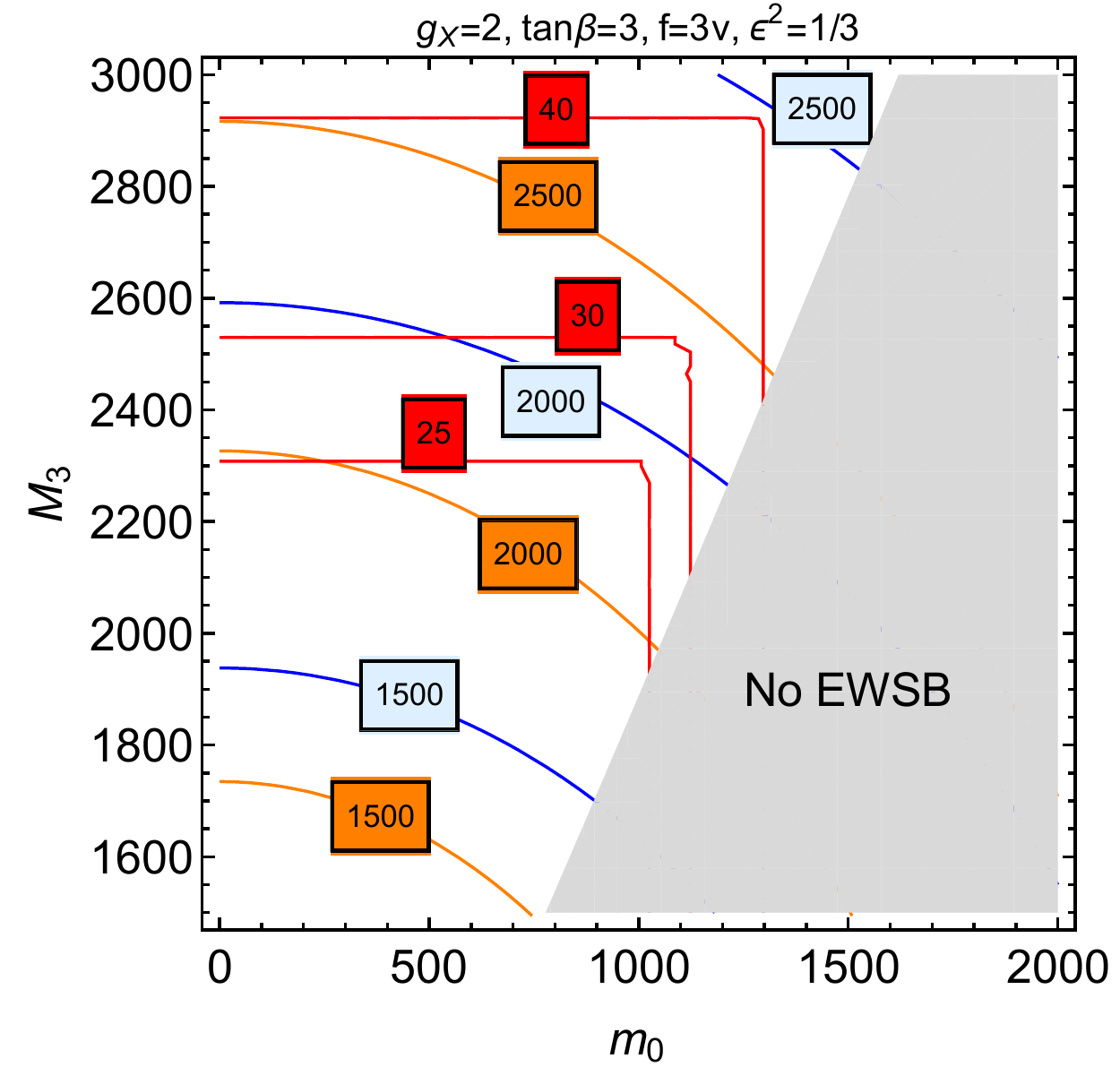}
 \caption{Masses of the lightest (right-handed) stop and the right-handed up squark (blue) and the lightest squark of the first generation other than
the right-handed up squark (orange) as a function of gluino soft mass parameter $M_3$
defined at the
stop mass scale  and universal soft scalar mass $m_0$ at the UV scale of $\Lambda=10^{16}$~GeV for $g_X=2$ and $\epsilon^2=1/3$.
}
\label{fig:EWSB}
\end{figure}

Using the above assumptions we show in Fig.~\ref{fig:EWSB} the contours of masses for the lightest stop 
and the
lightest first-generation squark other than the right-handed up squark in the plane $m_0$-$M_3$. The lightest stop is mostly right-handed and roughly
degenerate with the right-handed up squark. An important
constraint on the parameter space is provided by the condition of correct EWSB since the top yukawa coupling is much smaller
during the RG evolution
than in MSSM so the negative corrections from stops and gluino to $m_{H_u}^2$~\cite{Ibanez:1982fr,Inoue:1982pi} are smaller. This suppression is only partly compensated by the
negative correction from the $S$ fields. In consequence, for too large $m_0$, $m_{H_u}^2$ is positive at the low
scale. This can be easily circumvented by assuming $m_{H_u}$ smaller than $m_0$ at the mediation scale. Even without this assumption there are
parts of parameter space that give EWSB as well as viable sparticle spectrum.
In this example gluino is slightly heavier than squarks and the tuning at the
level of 5\% can be achieved with this simple UV boundary condition leading to squarks and gluino masses that comfortably satisfy the LHC
constraints. 
It may be  possible to reduce tuning even more if there exist some correlations between the soft
SUSY
breaking parameters in the UV leading to a focus point~\cite{Feng:1999mn} in which the overall correction to $m_{H_u}^2$ is small. 

{\it Flavor and collider phenomenology}---%
We should emphasize that asymptotic freedom for $g_X$ is obtained thanks to a small number of SM fermions charged under $SU(2)_X$. This implies
non-trivial flavor structure of the model which may  impact  flavor observables.
As explained before, we have assumed a flavor structure in yukawa couplings to suppress most of the tree-level FCNCs.
Tree-level FCNCs are, however, unavoidable in the top sector as we embed the right-handed up quark in $\bar{Q}_R$.

The heavy Higgs in ${\cal H}$ which we call $H_2$ couples to quarks via ${\cal L} = y_t H_2 \bar{u}_R Q_3$.
We expect non-negligible $t\to h u$ decays through mixing between the SM-like Higgs $h$ and the neutral component in $H_2$ which we call $H_2^0$.
The resultant $h$-$t$-$u$ coupling $\lambda_{htu}$ is as large as $m_Z^2 / m_{H_2}^2$.
The current upper limit on BR$(t\to h u)$ is $2\times10^{-3}$ corresponding to $\lambda_{htu}$ of about
$0.1$~\cite{Aaboud:2017mfd,Chiang:2015cba}, which implies a lower bound on $m_{H_2}$ of few hundred GeV.
The future sensitivity of the High-Luminosity LHC to BR$(t\to h u)$ is around $10^{-4}$~\cite{thu_HLLHC} so this process
will serve as an important probe of the model.

Flavor violation in the top sector has also impact on the rare decays of mesons.
We find that the strongest constraint comes from a possible deviation in ${\rm BR}(b\to s \gamma)$ due to one-loop corrections involving the charged
component of $H_2$, that we refer to as $H_2^\pm$, and the up quark, which is not suppressed by the GIM mechanism~\cite{GIM}.
Translating the bound obtained in~\cite{Misiak:2017bgg} for type-II Two-Higgs-Doublet model using the loop function in~\cite{Buras:1993xp}, we obtain
the lower bound $m_{H_2} \gtrsim$ 200 GeV.

The heavy Higgs $H_2$ is produced in proton colliders via the process $u +g \rightarrow H_2^+ b,~H_2^0 t$ involving the strong interaction and the top
yukawa coupling, 
with a dominant decay mode $H_2^+ \rightarrow u \bar{b}$, $H_2^0 \rightarrow u \bar{t},\bar{u}t$. None of existing searches give relevant
constraints on the masses of these new Higgses.

The right-handed up squark is almost degenerate in mass with the right-handed stop and decays mainly to the top/bottom quark and a higgsino. 
The signal resembles the one of the right-handed stop but with a much larger cross-section.

{\it Discussion.}---%
We have presented the first SUSY model which accommodates tuning of the EW scale $5-10\%$ for stops and gluino heavier than 2 TeV, 
satisfying current LHC constraints, even if
SUSY breaking mediation occurs close to the Planck scale. This is achieved in the novel UV completion of TH mechanism which is at work thanks to a new
asymptotically free $SU(2)_X$ gauge interaction. The model predicts that the right-handed up squark is degenerate with the right-handed stop and has 
a very different decay pattern than in the MSSM. This scenario 
may be also tested via flavor-violating top decays which are generically correlated with deviations from the SM prediction for the $b\to s \gamma$
decay. 

\section*{Acknowledgments}
This work has been partially supported by National Science Centre, Poland, under research grant DEC-2014/15/B/ST2/02157, by
the Office of High Energy Physics of the U.S. Department of Energy
under Contract DE-AC02-05CH11231, and by the National Science Foundation
under grant PHY-1316783. MB acknowledges support from the
Polish 
Ministry of Science and Higher Education through its programme Mobility Plus (decision no.\ 1266/MOB/IV/2015/0).


\begin{thebibliography}{99}

\bibitem{MaianiLecture}
L.~Maiani. in Proceedings: Summer School on Particle Physics, Paris,
	France (1979).
\bibitem{Veltman:1980mj} 
  M.~J.~G.~Veltman,
  Acta Phys.\ Polon.\ B {\bf 12}, 437 (1981).
\bibitem{Witten:1981nf} 
  E.~Witten,
  Nucl.\ Phys.\ B {\bf 188}, 513 (1981).
\bibitem{Kaul:1981wp} 
  R.~K.~Kaul,
  Phys.\ Lett.\ B {\bf 109}, 19 (1982).


\bibitem{Kaplan:1983fs} 
  D.~B.~Kaplan and H.~Georgi,
  Phys.\ Lett.\  {\bf 136B}, 183 (1984).

\bibitem{Kaplan:1983sm} 
  D.~B.~Kaplan, H.~Georgi and S.~Dimopoulos,
  Phys.\ Lett.\  {\bf 136B}, 187 (1984).


\bibitem{Chacko:2005pe}
  Z.~Chacko, H.~S.~Goh and R.~Harnik,
  Phys.\ Rev.\ Lett.\  {\bf 96} (2006) 231802
  [hep-ph/0506256].

\bibitem{Batra:2008jy}
  P.~Batra and Z.~Chacko,
  Phys.\ Rev.\ D {\bf 79} (2009) 095012
  [arXiv:0811.0394 [hep-ph]].

\bibitem{Geller:2014kta} 
  M.~Geller and O.~Telem,
  Phys.\ Rev.\ Lett.\  {\bf 114}, 191801 (2015)
  [arXiv:1411.2974 [hep-ph]].
  
  \bibitem{Barbieri:2015lqa}
  R.~Barbieri, D.~Greco, R.~Rattazzi and A.~Wulzer,
  JHEP {\bf 1508} (2015) 161
  [arXiv:1501.07803 [hep-ph]].
  
  \bibitem{Low:2015nqa}
  M.~Low, A.~Tesi and L.~T.~Wang,
  Phys.\ Rev.\ D {\bf 91} (2015) 095012
  [arXiv:1501.07890 [hep-ph]].
  
  
  \bibitem{Falkowski:2006qq}
  A.~Falkowski, S.~Pokorski and M.~Schmaltz,
  Phys.\ Rev.\ D {\bf 74} (2006) 035003
  [hep-ph/0604066].


\bibitem{Chang:2006ra}
  S.~Chang, L.~J.~Hall and N.~Weiner,
  Phys.\ Rev.\ D {\bf 75} (2007) 035009
  [hep-ph/0604076].
  
\bibitem{Craig:2013fga}
  N.~Craig and K.~Howe,
  JHEP {\bf 1403} (2014) 140
  [arXiv:1312.1341 [hep-ph]].

\bibitem{Katz:2016wtw}
  A.~Katz, A.~Mariotti, S.~Pokorski, D.~Redigolo and R.~Ziegler,
  JHEP {\bf 1701} (2017) 142
  [arXiv:1611.08615 [hep-ph]].


  
\bibitem{Badziak:2017syq}
  M.~Badziak and K.~Harigaya,
  JHEP {\bf 1706} (2017) 065
  [arXiv:1703.02122 [hep-ph]].

\bibitem{Badziak:2017kjk}
  M.~Badziak and K.~Harigaya,
  JHEP {\bf 1710} (2017) 109
  [arXiv:1707.09071 [hep-ph]].
  

  
\bibitem{Dimopoulos:2014aua}
  S.~Dimopoulos, K.~Howe and J.~March-Russell,
  Phys.\ Rev.\ Lett.\  {\bf 113} (2014) 111802
  [arXiv:1404.7554 [hep-ph]].

\bibitem{Batra:2003nj}
  P.~Batra, A.~Delgado, D.~E.~Kaplan and T.~M.~P.~Tait,
  JHEP {\bf 0402} (2004) 043
  [hep-ph/0309149].
  
\bibitem{Novikov:1983uc}
  V.~A.~Novikov, M.~A.~Shifman, A.~I.~Vainshtein and V.~I.~Zakharov,
  Nucl.\ Phys.\ B {\bf 229} (1983) 381.

\bibitem{Higgscomb}
  G.~Aad {\it et al.} [ATLAS and CMS Collaborations],
  JHEP {\bf 1608} (2016) 045
  [arXiv:1606.02266 [hep-ex]].

\bibitem{Buttazzo:2015bka}
  D.~Buttazzo, F.~Sala and A.~Tesi,
  JHEP {\bf 1511} (2015) 158
  [arXiv:1505.05488 [hep-ph]].

\bibitem{Ibanez:1982fr} 
  L.~E.~Ibanez and G.~G.~Ross,
  Phys.\ Lett.\  {\bf 110B}, 215 (1982).

\bibitem{Inoue:1982pi} 
  K.~Inoue, A.~Kakuto, H.~Komatsu and S.~Takeshita,
  Prog.\ Theor.\ Phys.\  {\bf 68}, 927 (1982)
  Erratum: [Prog.\ Theor.\ Phys.\  {\bf 70}, 330 (1983)].
  
\bibitem{Feng:1999mn}
  J.~L.~Feng, K.~T.~Matchev and T.~Moroi,
  Phys.\ Rev.\ Lett.\  {\bf 84} (2000) 2322
  [hep-ph/9908309].
 

 


\bibitem{Aaboud:2017mfd}
  M.~Aaboud {\it et al.} [ATLAS Collaboration],
  JHEP {\bf 1710} (2017) 129
  [arXiv:1707.01404 [hep-ex]].
  
\bibitem{Chiang:2015cba}
  C.~W.~Chiang, H.~Fukuda, M.~Takeuchi and T.~T.~Yanagida,
  JHEP {\bf 1511} (2015) 057
  [arXiv:1507.04354 [hep-ph]].
  
\bibitem{thu_HLLHC}  
ATLAS Collaboration, ATL-PHYS-PUB-2016-019.  
  
\bibitem{GIM}
  S.~L.~Glashow, J.~Iliopoulos and L.~Maiani,
  Phys.\ Rev.\ D {\bf 2} (1970) 1285.
  
\bibitem{Misiak:2017bgg}
  M.~Misiak and M.~Steinhauser,
  Eur.\ Phys.\ J.\ C {\bf 77} (2017) no.3,  201
  [arXiv:1702.04571 [hep-ph]].

\bibitem{Buras:1993xp}
  A.~J.~Buras, M.~Misiak, M.~Munz and S.~Pokorski,
  Nucl.\ Phys.\ B {\bf 424} (1994) 374
  [hep-ph/9311345].

\end{thebibliography}
\end{document}